\pdfoutput=1
\documentclass[final,twocolumn,11pt]{article}
\usepackage[utf8]{inputenc}
\usepackage[T1]{fontenc}
\usepackage{lmodern}
\usepackage[british]{babel}
\usepackage{csquotes}
\usepackage{amsthm}
\usepackage{authblk}
\usepackage[letterpaper,left=1.5cm, width=18.5cm, top=2cm, bottom=2cm]{geometry}
\usepackage{abstract}
\addto\captionsenglish{\renewcommand{\abstractname}{}}   %
\renewcommand{\absnamepos}{empty} %
\usepackage{titlesec}
\titleformat*{\section}{\large\bfseries}
\titleformat*{\subsection}{\normalsize\mdseries}
\usepackage[
redefine-symbols=false,
  input-uncertainty-signs=\pm,
  separate-uncertainty = true,
  group-separator = {\,},
	detect-weight=true,
  detect-family,
]{siunitx}
\usepackage[
  backend=biber,
  citestyle=authoryear-comp,
  bibstyle=authoryear,
  natbib=true,
  sortlocale=auto,
  sorting=nyt,
  urldate=iso,
  date=iso,
  seconds=true,
  sortcites=false,
  url=false,
  doi=true,
  mincitenames=1, maxcitenames=2,	maxbibnames=99,
  eprint=true, 
  mincrossrefs=2,minxrefs=2,
  useprefix=true,
  giveninits=true,
  uniquename=init,
  uniquelist=minyear
]{biblatex}
\renewcommand*{\bibfont}{\footnotesize}
\usepackage[pdftex]{graphicx} %
\usepackage{pgfplots}
\usepackage{xcolor}
\pgfplotsset{compat=1.14}
\usepgfplotslibrary{fillbetween}
\usepackage{epstopdf}
\usepackage{subcaption} %
\usepackage{hyperref}
\usepackage{nicefrac}
\usepackage{amsmath}
\usepackage{tikz}
\usetikzlibrary{plotmarks}
\usepackage{xspace}
\newbox{\myorcidaffilbox}
\sbox{\myorcidaffilbox}{\large\includegraphics[height=1.7ex]{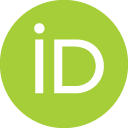}}
\newcommand{\orcid}[1]{\href{https://orcid.org/#1}{\usebox{\myorcidaffilbox}}}
\DeclareSIUnit[per-mode=symbol]\nms{\nm\per\s\squared}
\DeclareSIUnit\nmsi{\nm\per\s\squared}
\DeclareSIUnit[per-mode=symbol]\mps{\m\per\s\squared}
\DeclareSIUnit[]\mhz{\mega\hertz}
\DeclareSIUnit[]\month{month}
\newcommand{\density}[1]{\SI[per-mode=symbol]{#1}{\kg\per\cubic\m}}
\newcommand{\nmwert}[1]{\SI{#1}{\nms}}
\newcommand{\mwert}[1]{\SI{#1}{\m}}

\renewcommand{\vec}[1]{\mathbf{#1}}
\newcommand{\keff}{\ensuremath{k_\mathrm{eff}}\xspace}
\newcommand{\zeff}{\ensuremath{z_\mathrm{eff}}\xspace}
\newcommand{\vkeff}{\vec{k}_\ensuremath{\mathrm{eff}}\xspace}
\newcommand{\textg}{\textit{g}\xspace}
\newcommand{\eg}{e.\,g.\@\xspace}
\theoremstyle{definition}
\newtheorem*{authorcontrib}{Author contributions}
\newtheorem*{dataavail}{Data availability statement}
\newtheorem*{acknowledgements}{Acknowledgements}
\colorlet{mygray}{black!50}
\colorlet{myblue}{blue!50}
\colorlet{myred}{red}
\colorlet{mygreen}{green!75}

\DeclareRobustCommand\showmark[2]{\tikz[]{\node[#2]{\pgfuseplotmark{#1}};}}

\colorlet{c-roi}{blue!50!white}
\colorlet{c-baseline}{green!40!gray}
\colorlet{c-sas}{magenta!80!gray}
\colorlet{c-sources}{red!60!gray}
\DeclareRobustCommand\showline[2]{\tikz[baseline=-0.5ex]{\draw[#1,line width=#2] (0,0) -- (0.3,0);}}

\begin{document}
\renewcommand{\abstractname}{}
\renewcommand{\absnamepos}{empty} %
\setlength{\abstitleskip}{-0.8cm}
\title{Gravity field modelling for the Hannover \SI{10}{\m} atom interferometer}

\author[1,2]{Manuel~Schilling\orcid{0000-0002-9677-0119}}
\author[3]{\'Etienne~Wodey\orcid{0000-0001-9522-8558}}
\author[2]{Ludger~Timmen\orcid{0000-0003-2334-5282}}
\author[3]{Dorothee~Tell\orcid{0000-0002-2793-7982}}
\author[3]{Klaus~H.~Zipfel\orcid{0000-0002-3135-1247}}
\author[3]{Dennis~Schlippert\orcid{0000-0003-2168-1776}}
\author[1,3]{Christian~Schubert}
\author[3]{Ernst~M.~Rasel}
\author[2]{J\"urgen~M\"uller\orcid{0000-0003-1247-9525}}

\affil[1]{German Aerospace Center (DLR), Institute for Satellite Geodesy and Inertial Sensing, c/o Leibniz Universit\"at Hannover, DLR-Institut, Welfengarten 1, 30167 Hannover, Germany}
\affil[2]{Leibniz Universit\"at Hannover, Institut f\"ur Erdmessung, Schneiderberg 50, 30167 Hannover, Germany}
\affil[3]{Leibniz Universit\"at Hannover, Institut f\"ur Quantenoptik, Welfengarten 1, 30167 Hannover, Germany}
\affil[ ]{{\normalfont E-mail: \href{mailto:manuel.schilling@dlr.de}{\ttfamily manuel.schilling@dlr.de}}}
\renewcommand\Affilfont{\itshape\small}
\date{\footnotesize This is a post-peer-review, pre-copyedit version of an article published in Journal of Geodesy 94:122. The final authenticated version is available online at: \url{https://dx.doi.org/10.1007/s00190-020-01451-y}}
\twocolumn[
\maketitle
\begin{onecolabstract}  
Absolute gravimeters are used in geodesy, geophysics, and physics for a wide spectrum of applications. 
Stable gravimetric measurements over timescales from several days to decades are required to provide relevant insight into geophysical processes. 
Users of absolute gravimeters participate in comparisons with a metrological reference in order to monitor the temporal stability of the instruments and determine the bias to that reference. 
However, since no measurement standard of higher-order accuracy currently exists, users of absolute gravimeters participate in key comparisons led by the International Committee for Weights and Measures.
These comparisons provide the reference values of highest accuracy compared to the calibration against a single gravimeter operated at a metrological institute.
The construction of stationary, large scale atom interferometers paves the way towards a new measurement standard in absolute gravimetry used as a reference with a potential stability up to \nmwert{1} at \SI{1}{\second} integration time. 
At the Leibniz University Hannover, we are currently building such a very long baseline atom interferometer with a \SI{10}{\meter} long interaction zone. 
The knowledge of local gravity and its gradient along and around the baseline is required to establish the instrument's uncertainty budget and enable transfers of gravimetric measurements to nearby devices for comparison and calibration purposes. 
We therefore established a control network for relative gravimeters and repeatedly measured its connections during the construction of the atom interferometer. 
We additionally developed a 3D model of the host building to investigate the self-attraction effect and studied the impact of mass changes due to groundwater hydrology on the gravity field around the reference instrument.
The gravitational effect from the building 3D model is in excellent agreement with the latest gravimetric measurement campaign which opens the possibility to transfer gravity values with an uncertainty below the \nmwert{10} level.
\par

\vspace{0.2cm}
 \par
\end{onecolabstract}
\vspace{0.7cm}
]
\section{Introduction}
A variety of applications in geodesy, geophysics and physics require the knowledge of local gravity \textg \citep{VanCamp2017}. 
These applications include observing temporal variations of the mass distribution in the hydrosphere, atmosphere and cryosphere and furthermore the establishment and monitoring of height and gravity reference frames, the determination of glacial isostatic adjustment, and the realisation of SI\footnote{Syst\`eme International d'unit\'es} units, \eg, of force and mass \citep{Merlet2008,Liard2014,Schilling2017}.
The absolute value of gravity \textg is usually measured by tracking the free-fall of a test mass using a laser interferometer \citep{Niebauer1995}. 
The operation of an absolute gravimeter (AG), especially the combination of several instruments in a project, requires special consideration of the offset to \emph{true g} and the change thereof. 
In addition, the long-term stability of absolute gravimeters is of particular relevance when measuring small gravity trends. 
For example, the determination of the glacial isostatic adjustment (GIA) on regional scales of around \SI{1000}{\kilo\meter} \citep{Timmen2011} requires an instrument stable to the \nmwert{20} level over several years. 
Extending this effort by deploying several AGs also requires the knowledge of the biases of all the instruments involved \citep{Olsson2019}. 
The lack of a calibration service with a \SIrange{10}{20}{\nms} uncertainty requires the participation in key comparisons \citep[KC, \eg ][]{Falk2019} where the reference values are determined with an uncertainty of approximately \nmwert{10}. 
This uncertainty level requires the participation of multiple gravimeters and cannot be achieved by comparison against a single gravimeter operated at a metrological institute.
However, the development of stationary atom interferometers, which can be operated as gravimeters, so-called quantum gravimeters (QG), may result in such a superior reference in the future available for regular comparisons or on demand by the user.
A major requirement in this respect is the control of systematic effects like wavefront aberration or the Coriolis effect.
In this paper, we focus on the modelling and measurement of the local gravity field.

We start by discussing the typical approaches for monitoring the long-term stability of an AG and tracing the measurements back to the SI (section~\ref{sec:trace}). 
Then, after briefly describing the working principle of atomic gravimeters and the case for very long baseline atom interferometry (section~\ref{sec:vlbag}), we present a gravity model for the Hannover Very Long Baseline Atom Interferometry (Hannover-VLBAI) facility, a new \mwert{10}-scale baseline atom interferometer in commissioning at the Leibniz University Hannover (section~\ref{sec:envmod}). 
Finally, we present the micro-gravimetric surveys performed at the instrument's site (section~\ref{sec:gravmeas}) to assess the accuracy of the gravity model (section~\ref{sec:transg}). 
This paves the way towards control of the systematics in the atom interferometer and accurate transfers of measured \textg values between the VLBAI operating as a gravimeter and transportable AGs in a nearby laboratory.
\section{Gravimeter bias and SI traceability}
\label{sec:trace}
Micro-g LaCoste FG5(X) \citep{Niebauer2013} instruments represent the current state of the art in absolute gravimetry. 
They track the trajectories of a free-falling test mass with corner cubes by means of laser interferometry to determine the local acceleration of gravity \textg. 
These types of absolute gravimeters are referred to as \emph{classical absolute gravimeters} in the following text.

As described by the 2015 CCM-IAG\footnote{Consultative Committee for Mass and related quantities -- International Association of Geodesy} Strategy for Metrology in Absolute Gravimetry \citep{CCM2015}, there are two complementary paths for the traceability of absolute gravity measurements: a) calibration of incorporated frequency generators and b) additional gravimeter comparisons against a reference. 
The direct way of tracing absolute gravity measurements back to the SI goes through the calibration of their incorporated laser and oscillator to standards of length and time \citep{Vitushkin2011}. 
In high-accuracy instruments, the laser frequency is typically locked to a standard transition of molecular iodine \citep{Chartier1993, Riehle2018}. 
The time reference is usually given by a rubidium oscillator which needs to be regularly compared with a reference oscillator to ensure its accuracy as external higher-accuracy time sources are typically not available at measurement sites. 
In most cases, the oscillator's frequency drift is linear $(\SI[per-mode=symbol]{<0.5}{\milli\hertz\per\month} \text{ or } \SI[]{<1}{\nm\per\s\squared}/\si{\month})$ and a few calibrations per year are sufficient.
However, \citet{Maekinen2015} and \citet{Schilling2016} report on sudden jumps in frequency\footnote{Current publications refer to the Microsemi (formerly Symmetricon) SA.22c rubidium oscillator} equivalent to several tens of \si{\nms} due to increased concentrations of gaseous helium \citep{Riehle2004} when measuring near superconducting gravimeters. 
Such higher concentrations might occur after installation, maintenance, or repair of a superconducting gravimeter and are unlikely during normal operation.
The frequency drift changes to an exponential decrease after the helium event and may remain this way for years \citep{Schilling2019}. 

\begin{figure}%
\includegraphics[width=\columnwidth]{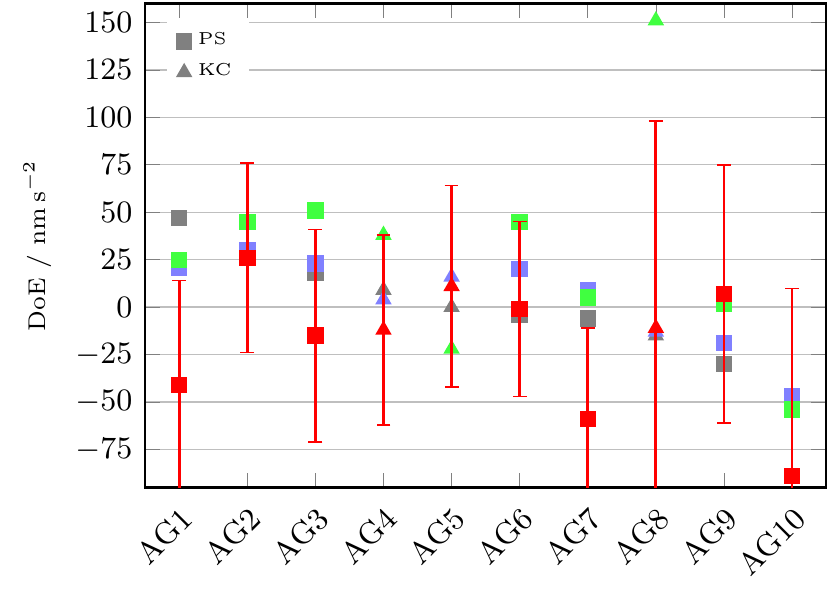}%
\caption{Degree of Equivalence (DoE) of joint participants of EURAMET.M.G-K1 \citep[\showmark{square*}{mygray}][]{Francis2013}, CCM.G-K2 \citep[\showmark{square*}{myblue}][]{Francis2015}, EURAMET.M.G-K2 \citep[\showmark{square*}{mygreen}][]{Palinkas2017} and EURAMET.M.G-K3 \citep[\showmark{square*}{myred}][]{Falk2019}. The participants are sorted by DoE of the first KC. The expanded uncertainty is given only for the last KC. Pilot Study (PS) indicates instruments of non NMI/DI institutions. All AGs shown are laser interferometers of which eight are FG5(X) type instruments.}%
\label{fig:agvgl}%
\end{figure}

The equivalence of gravity measurement standards and the definition of the gravity reference are established by international comparisons in the framework of the CIPM MRA\footnote{Mutual Recognition Agreement of the Comit\'e International des Poids et Mesures}. 
Since no higher-order reference instrument is available, key comparisons are held in an approximately two-year interval, alternating between CIPM key comparisons and regional comparisons.  
There, the instruments operated by National Metrology Institutes (NMI) and Designated Institutes (DI) are used to determine the Key Comparison Reference Value (KCRV). 
The bias to the KCRV, or Degree of Equivalence (DoE) is then calculated for all individual instruments, including those without NMI/DI status participating in the so-called pilot study (PS), and serves as validation for their uncertainty.

Figure~\ref{fig:agvgl} shows the common participants, out of a total number of 35 gravimeters participating in the comparisons, to the last four KC held in Europe \citep{Francis2013,Francis2015,Palinkas2017,Falk2019}. 
One observes that the spread of DoE over all instruments is around \nmwert{\pm 75}, and at a similar level for the most extreme cases of individual instruments. 
Even though the DoEs of the instruments in these comparisons are typically within the uncertainties declared by the participants, figure \ref{fig:agvgl} also shows the necessity of determining these biases of gravimeters, classical and quantum alike, to monitor an instrument's stability in time. 
Biases can then be taken into account in gravimetric projects. 
The variation of the bias of an instrument can be explained by a variety of factors. 
For example, \citet{Olsson2016} show that a permanent change in the bias of a classical AG can occur during manufacturer service or unusual transport conditions (\eg aviation transport). 
Also, \citet{Kren2017,Kren2019} identified, characterised and partially removed biases originating in the signal processing chain of FG5 gravimeters, \eg due to cable length and fringe signal amplitude.

Regional KCs are linked to a CIPM KC by a small number of common NMI/DI participants applying the so-called linking converter \citep[typically around \nmwert{\pm 10},][]{Jiang2013}. 
The underlying assumption is that instrumental biases of the NMI/DI instruments remain stable \citep{Delahaye2002}. 
Otherwise, this would introduce an additional shift in the bias of all participating instruments of the regional KC and PS. 

Quantum gravimeters, based on matter wave interferometry with cold atoms, offer a fully independent design. 
They have demonstrated stabilities and accuracies at levels comparable to those from state of the art classical AGs by participating in KCs \citep{Gillot2016,Karcher2018} or common surveys with other instruments at various locations \citep{Freier2017,Schilling2019}. 
The availability of improved QGs as gravity references provides an opportunity to enhance the stability of reference values obtained during key comparisons and therefore lead to an international gravity datum of better stability in time.
Just by that alone QGs could become a serious alternative to classical absolute gravimeters.
\section{Very long baseline atomic gravimetry}
\label{sec:vlbag}
\subsection{Atominterferometric gravimetry}
\label{sec:atomic_gravimetry}

Most atomic gravimeters use cold matter waves as free-falling test masses to measure absolute gravity. 
They exploit the coherent manipulation of the external degrees of freedom of these atomic test masses with light pulses to realise interferometers sensitive to inertial quantities and other forces. 
These techniques are for example used to perform precision measurements of fundamental constants \citep{Rosi2014,Bouchendira2011,Parker2018}, test fundamental physics \citep{Schlippert2014,Rosi2017,Jaffe2017}, sense small forces \citep{Alauze2018} and perform gravimetry, gravity-gradiometry, and measure rotations with record instabilities and inaccuracies \citep{Menoret2018,Freier2016,Gillot2014,Zhou2012,Savoie2018,Sorrentino2014}.

\begin{figure}
  \centering
  \includegraphics[width=\columnwidth]{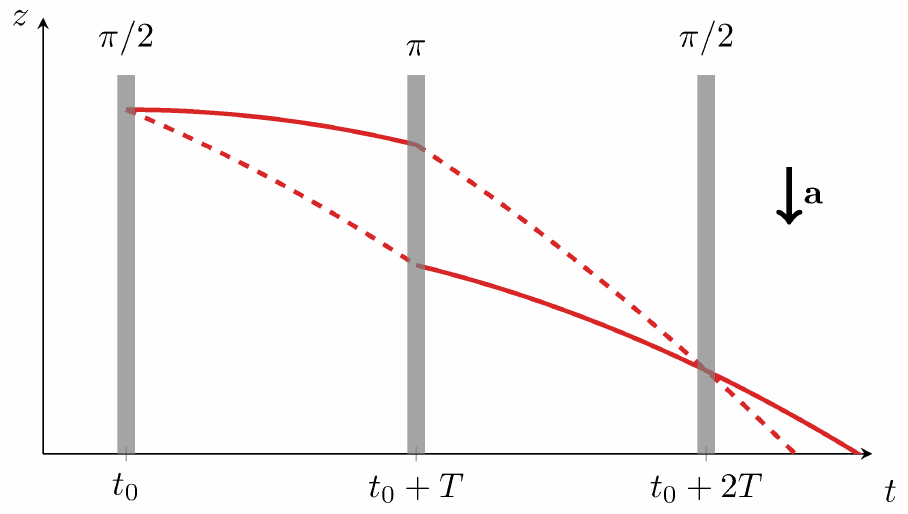}
  \caption{Mach--Zehnder light-pulse atom interferometer geometry in a uniform acceleration field $\vec{a}$. At time $t_0$, the atomic matterwave is put in a superposition of momenta $p$ (\showline{red!70!gray}{0.4mm}) and $p+\hbar\keff$ (\showline{red!70!gray,dashed}{0.4mm}). The momenta are reversed at time $t_0+T$ to recombine the wave packets with a last light pulse at time $t_0+2T$. The populations in the two momentum classes after the last light pulse allow extracting the interferometric phase $\Delta\phi$.}
  \label{fig:ai-mz-geometry}
\end{figure}

Atomic gravimeters typically realise the Mach--Zehnder light-pulse atom interferometer geometry \citep{Kasevich1991} depicted in figure~\ref{fig:ai-mz-geometry}. 
In this analogon to the eponymous configuration for optical interferometers, the leading-order interferometric phase $\Delta\phi$ scales with the space-time area enclosed by the interferometer:

\begin{equation}
  \label{eq:ai-mz-phase}
  \Delta\phi = \vkeff\cdot\vec{a} T^2
\end{equation}

where $\hbar\vkeff$ is the recoil transfered to the atomic wave packets by the atom-light interaction processes (cf. figure \ref{fig:ai-mz-geometry}, $\hbar$ is the reduced Planck constant and $\vkeff$ the effective optical wave vector), $\vec{a}$ the uniform acceleration experienced by the atoms during the interferometric sequence, and $T$ the pulse separation time.
The full interferometer has a duration of $2T$.
The knowledge of the instrument's scale factor $\keff T^2$ and the measurement of the phase $\Delta\phi$ allow determining the projection of the acceleration $\vec{a}$ along $\vkeff$. 
When $\vkeff$ is parallel to $\vec{g}$, such an instrument can therefore be used as a gravimeter, measuring the total vertical acceleration of the matter waves used as test masses.

The Mach--Zehnder light-pulse atom interferometer works as follows. 
For each interferometric sequence, a sample of cold atoms is prepared in a time $T_p$. 
Then, at time $t=t_0$, the first atom-light interaction pulse puts the matter wave in a superposition of quantum states with different momenta $\vec{p}$ and $\vec{p}+\hbar\vkeff$, thus effectively creating two distinct semi-classical trajectories. 
At time $t=t_0+T$, a second atom-light interaction process redirects the two atomic trajectories to allow closing the interferometer at time $t=t_0+2T$ with a third light pulse. 
Counting the population of atoms in the two momentum states provides an estimation of the interferometric phase $\Delta\phi$. 
Finally, the cycle of preparation of the cold atoms, coherent manipulation of the matter waves, and detection is repeated. 
Since the atom-light interaction imprints the local phase of the light on the matter waves, the above measurement principle can be interpreted as measuring the successive positions of a free-falling matter wave at known times $t_0$, $t_0+T$, and $t_0+2T$ with respect to the light field. 
The inertial reference frame for the measurement system, similar to the superspring in FG5(X) gravimeters, is usually realised by a mirror retro-reflecting the light pulses, creating well-defined equiphase fronts.
Practically, the interferometric phase $\Delta\phi$ is scanned by accelerating the optical wave fronts at a constant rate $\alpha$, effectively continuously tuning the differential velocity between the matter waves and the optical equiphase fronts.
Assuming that $\vkeff$ and $\vec{a}$ are parallel, the interferometric phase reads:
\begin{equation}
  \label{eq:ai-mz-phase-alpha}
  \Delta\phi = \keff \left(a - \frac{\alpha}{\keff}\right) T^2\ .
\end{equation}
When $\alpha = \keff a$, the interferometric phase vanishes independently of the interferometer's duration $2T$, allowing to unambiguously identify this operation point. 
Physically, $\alpha = \keff a$ exactly compensates the Doppler effect experienced by the atomic matter waves due to the acceleration $a$.
Therefore, the measurement of the acceleration $a$ amounts to a measurement of the acceleration rate $\alpha$ which can be traced back to the SI since it corresponds to frequency generation in the radio-frequency domain.

Assuming white noise at a level $\delta\phi$ for the detection of the interferometric phase, the instrument's instability is given by:
\begin{equation}
  \label{eq:ai-short-term-instability}
  \delta a(\tau) = \sqrt{2T+T_p}\cdot\frac{\delta\phi}{\keff T^2}\cdot\frac{1}{\sqrt{\tau}}\ .
\end{equation}
where $\tau$ is the measurement's integration time.
This expression reveals the three levers for reducing the measurement instability: decreasing the single shot noise level $\delta\phi$, increasing the scale factor $\keff T^2$, and minimising the sample preparation time $T_p$, as it contributes to the total cycle time without providing phase information.
In transportable devices, record instabilities have been achieved by \citet{Freier2016} with $\delta a = \nmwert{96}$ at $\tau = \SI{1}{\s}$. Commercial instruments like the Muquans AQG \citep{Menoret2018} reached instabilities of \nmwert{500} at $\tau = \SI{1}{\s}$ with sample rates up to \SI{2}{\hertz}. 
The dominant noise source is vibrations of the mirror realising the reference frame for the measurements.

The accuracy of such quantum gravimeters stems from the well-controlled interaction between the test masses and their environment during the measurement sequence. 
The main sources of inaccuracy in such instruments originate from uncertainties in the atom-light interaction parameters (\eg imperfections of the equiphase fronts of the light wave), stray electromagnetic field gradients creating spurious forces, thus breaking the free-fall assumption, and knowledge of the inhomogeneous gravity field along the trajectories. 
Extensive characterisation of these effects led to uncertainties in QGs below \nmwert{40}, consistent with the results from CIPM key comparisons \citep{Gillot2014} or common surveys with classical AGs \citep{Freier2016}.

\subsection{Very Long Baseline Atom Interferometry}
\label{sec:vlbai}

Very Long Baseline Atom Interferometry (VLBAI) represents a new class of ground-based atom interferometric platforms which extends the length of the interferometer\textquotesingle s baseline from tens of centimetres like in typical transportable instruments \citep{Freier2016,Gillot2014} to multiple meters. 
According to equation~\eqref{eq:ai-mz-phase}, the vertical acceleration sensitivity of a Mach--Zehnder type atom interferometer scales linearly with the length of the baseline ($\sim aT^2$). 
Therefore, an increase in the length of the baseline potentially enables a finer sensitivity for the atomic gravimeter through an increased scale factor $\keff T^2$. 
A \SI{10}{\meter}-long baseline instrument can for example extend the interferometric time $2T$ to around \SI{1}{\second} if the atoms are simply dropped along the baseline or up to \SI{2.4}{\second} if they are launched upwards in a fountain-like fashion.
In the simple drop case, the velocity acquired by the atoms between their release from the source and the start of the interferometer leads to an interferometer duration shorter than half of the one for the launch case.
For our apparatus, the distance between the top source chamber and the region of interest is around \SI{2}{\meter} (see figure~\ref{fig:vlbai}), constraining $T < \SI{400}{\milli\second}$ for simple drops.

Using realistic parameters ($T_p=\SI{3}{\s}$, $\delta\phi=\SI{10}{\milli\radian}$), equation~\eqref{eq:ai-short-term-instability} yields potential short-term instabilities for VLBAIs ($\tau = \SI{1}{\second}$ integration time):
\begin{equation}
  \label{eq:ai-short-term-stability-vlbai}
  \begin{array}{ll}
    T = \SI{400}{\milli\second}\text{:~} & \delta a = \nmwert{8} \\
    T = \SI{1.2}{\second}\text{:~} & \delta a = \nmwert{1}
  \end{array}
\end{equation}
competing with the noise level of superconducting gravimeters \citep{Rosat2011,Rosat2018} while providing absolute values of the gravity acceleration \textg.

Nevertheless, the increased scale factor $\keff T^2$ gained by the expanded baseline comes at the price of a stationary device with added complexity due to its size, and a vibration noise sensitivity magnified by the same scale factor as the gravitational acceleration for frequencies below $\nicefrac{1}{(2T)}$. 
Hence, the use of VLBAIs as ultra stable gravimeters requires new developments in the control of environmental vibrations \citep{Hardman2016}. 
Also, time- and space-varying electromagnetic and gravity fields along the free-fall trajectories of the matter waves have a direct impact on the accuracy and stability of the instrument, as the corresponding spurious forces depart from the assumptions of equation~\eqref{eq:ai-mz-phase}, therefore leading to biases \citep{DAgostino2011} and impacting the instrument's effective height \citep{Timmen2003}.

\subsection{Effective height}
\label{subsec:effh}
In order to compare measurements of a VLBAI gravimeter with other instruments, it is crucial to determine the effective height $\zeff$ defined by:

\begin{equation}
  g_0 - \gamma\zeff = \frac{\Delta\phi_\mathrm{tot}}{\keff T^2}
\end{equation}
where $g_0\approx\SI{9.81}{\meter\per\second}$ is the value of gravity at $z=0$, $\gamma\approx\SI{3}{\micro\meter\per\second\squared\per\meter}$ the magnitude of the linear gravity gradient, and $\Delta\phi_\mathrm{tot}$ the phase shift measured by the interferometer.
The right-hand side is the value of gravity measured by the atom interferometer, including all bias sources.
Restricting to first order in the gravity-gradient $\gamma$, and applying a path-integral formalism, one gets \citep{Peters2001}:

\begin{equation}
  \label{eq:def-zeff}
  \zeff = z_0 - \dfrac{\Delta g}{\gamma}\quad\text{with}\quad\Delta g = \frac{7}{12}\gamma g_0 T^2 - \gamma \bar{v}_0T
\end{equation}
where $z_0$ is the height of the start of the interferometer and $\bar{v}_0 = v_0 + \nicefrac{\hbar\keff}{(2m)}$ the mean atomic velocity just after the interferometer opens ($v_0$ is the atomic velocity before the first beamsplitter, and $m$ is the atomic mass).
This expression for \zeff is compatible with the one given for FG5 gravimeters by \cite{Palinkas2012}.
In particular, it only depends on the value of the gradient $\gamma$ through $v_0$ and $z_0$.
Indeed, the interferometer is controlled in time and the initial position and velocity $z_0$ and $v_0$ are therefore given by the free-fall motion of the atoms between the source chamber and the region of interest.
In general, \zeff depends on the geometry of the atom interferometer.
For the simple drop case in the Hannover VLBAI facility (see section~\ref{sec:hannover-vlbai}), $\zeff\approx\SI{9.2}{\meter}$.

Corrections to equation~\eqref{eq:def-zeff} must be taken into account to constrain the uncertainty on gravity at $\zeff$ below \SI{10}{\nano\meter\per\second\squared}.
On the one hand, terms of order $\gamma^2$ and higher in $\Delta\phi_\mathrm{tot}$ contribute at the sub-\si{\nano\meter\per\second\squared} level.
On the other hand, one can use perturbation theory \citep{Ufrecht2020} to estimate the effect of the non-homogeneous gravity gradient along the interferometer's baseline.
Using the data discussed here, we evaluate this effect below \SI{5}{\nano\meter\per\second\squared}, therefore lying within the model's uncertainty (see section~\ref{sec:transg}) and similar to the known contribution for FG5(X) gravimeters \citep{Timmen2003}.

Finally, when using multiple concurrent interferometers at different heights, the effect of a homogeneous gravity gradient can be mitigated by measuring it simultaneously with the acceleration value \citep{Caldani2019}.
In this case, the effective height corresponds to the position of the mirror giving the inertial reference.
Detailed modelling is however still necessary to push the uncertainty budget in the sub-\SI{10}{\nano\meter\per\second\squared} and calibrate the instrument to the level of its instability.

\subsection{The Hannover VLBAI facility}
\label{sec:hannover-vlbai}

We introduce the Hannover Very Long Baseline Atom Interferometry facility, an instrument developed at the newly founded Hannover Institute of Technology (HITec) of the Leibniz University Hannover, Germany. 
It builds on the concepts outlined in section~\ref{sec:vlbai} to provide a platform to tackle challenges in extended baseline atom interferometry. 
In the long term, it aims at tests of our physical laws and postulates like for example the universality of free fall \citep{Hartwig2015}, searches for new forces or phenomena, and the development of new methods for absolute gravimetry and gravity gradiometry \citep{Schlippert2019}.

\begin{figure}
{\small
  \centering
  \begin{tikzpicture}
  \begin{axis}[
    width=7.3cm, height=15.4cm,
    scale only axis,
    enlarge x limits={value=0.5, lower},
    xmax=6,
    axis on top, xtick=\empty, axis lines=left,
    x axis line style={draw=none},
    ylabel={Height / \si{\m}},
    ]
    \addplot graphics[xmin=-1.3,xmax=1.3,ymin=-0.06,ymax=15.4] {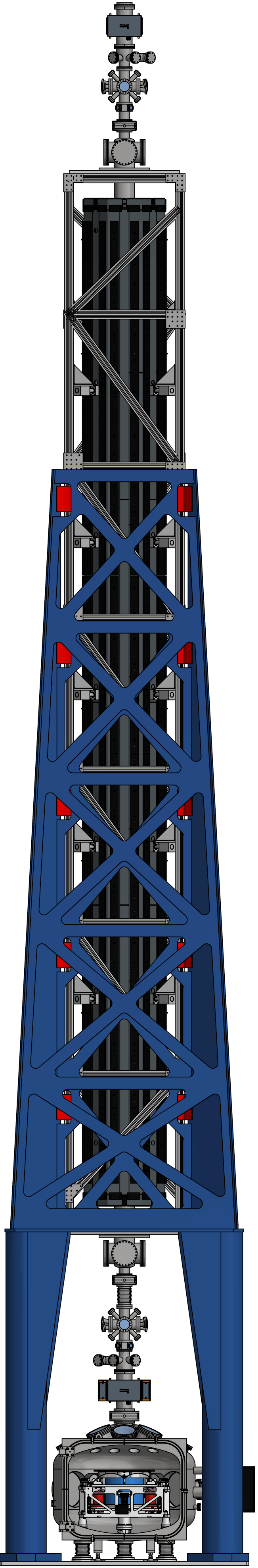};
    \begin{scope}[mark=none, domain=-1.3:1.3, draw=c-roi]
      \addplot+ [name path=roi-up] {12.5};
      \addplot+ [name path=roi-down] {4.3};
      \addplot [c-roi,opacity=0.4] fill between [of=roi-up and roi-down];
    \end{scope}
    \begin{scope}[rounded corners, line width=0.8mm]
      \draw[c-baseline] (axis cs:-0.8,14.1) rectangle (axis cs:0.7,2.9);
      \draw[c-sources] (axis cs:-0.8,2.1) rectangle (axis cs:0.7,2.6);
      \draw[c-sources] (axis cs:-0.8,14.3) rectangle (axis cs:0.7,14.8);
      \draw[c-sas] (axis cs:-0.8,1.4) rectangle (axis cs:0.7,0.1);
    \end{scope}
    \begin{scope}[every node/.style={align=left, anchor=west}]
      \node[c-sources] at (axis cs:1.6,14.5) {Upper \textbf{atomic source}};
      \node[c-sources] at (axis cs:1.6,2.4) {Lower \textbf{atomic source}};
      \node[c-baseline] at (axis cs:1.6,8.5) {\textbf{Baseline}\\ultra-high vacuum chamber\\and magnetic shield};
      \node[c-sas] at (axis cs:1.6,0.8) {\textbf{Inertial reference}\\vibration isolated mirror};
      \node[c-roi] at (axis cs:1.6,12.) {\textbf{Region of interest}\\for precision\\atom interferometry};
    \end{scope}
  \end{axis}
\end{tikzpicture} }
  \caption{The Hannover Very Long Baseline Atom Interferometry (VLBAI) facility and its three main elements: source chambers \showline{c-sources}{0.8mm}, baseline \showline{c-baseline}{0.8mm}, and inertial reference system and vacuum vessel (VTS) \showline{c-sas}{0.8mm}. The baseline and upper source chambers are supported by an aluminium structure (VSS, dark blue). The region of interest for atom interferometry is shaded in light blue.}
  \label{fig:vlbai}
\end{figure}

The Hannover VLBAI facility is built around three main elements shown in figure~\ref{fig:vlbai}:

\begin{enumerate}
\item Ultra cold samples of rubidium and ytterbium atoms are prepared in the two \emph{source chambers}, allowing for both drop (max $T=\SI{400}{\milli\second}$) and launch (max $T=\SI{1.2}{\s}$) modes of operation. Advanced atom-optics promise enhanced free-fall times by relaunching the wave packets during the interferometric sequence \citep{Abend2016};
\item The reference frame for the inertial measurements is realised by a \emph{seismically isolated mirror} at the bottom of the apparatus. The seismic attenuation system (SAS) uses geometric anti-spring filters \citep{Wanner2012} to achieve vibration isolation above its natural resonance frequency of \SI{320}{\milli\hertz}. The isolation platform is operated under high vacuum conditions to reduce acoustic and thermal coupling. The vacuum vessel containing the SAS is denoted VTS in sections~\ref{sec:envmod}--\ref{sec:transg};
\item The \mwert{10.5}-long \emph{baseline} consists of a \SI{20}{\centi\meter} diameter cylindrical aluminium vacuum chamber and a high-performance magnetic shield \citep{Wodey2019}. The interferometric sequences take place along this baseline, in the \mwert{8}-long central \emph{region of interest} where the longitudinal magnetic field gradients fall below \SI[per-mode=symbol]{2.5}{\nano\tesla\per\meter}.
\end{enumerate}

In order to decouple the instrument from oscillations of the walls of the building, the apparatus is only rigidly connected to the foundations of the building. 
The VTS (and SAS) and lower source chamber are mounted on a baseplate directly connected to the foundation. 
The baseline and upper source chamber are supported by a \mwert{10} high aluminium tower, denoted as VLBAI support structure (VSS) in the following sections. 
The footprint of the device on the floor is $\mwert{2.5}\times\mwert{2.5}$. 
Traceability to the SI is ensured by locking the instrument's frequency references to standards at the German NMI (PTB Braunschweig) via an optical link \citep{Raupach2015}. 
All heights are measured from the instrument's baseplate. 
The altitude of this reference point in the German height datum is \mwert{50.545}.

\section{Environmental model}
\label{sec:envmod}

\begin{figure*}
  \begin{subfigure}[b]{0.38\textwidth}
    \centering
    \includegraphics[height=7.5cm]{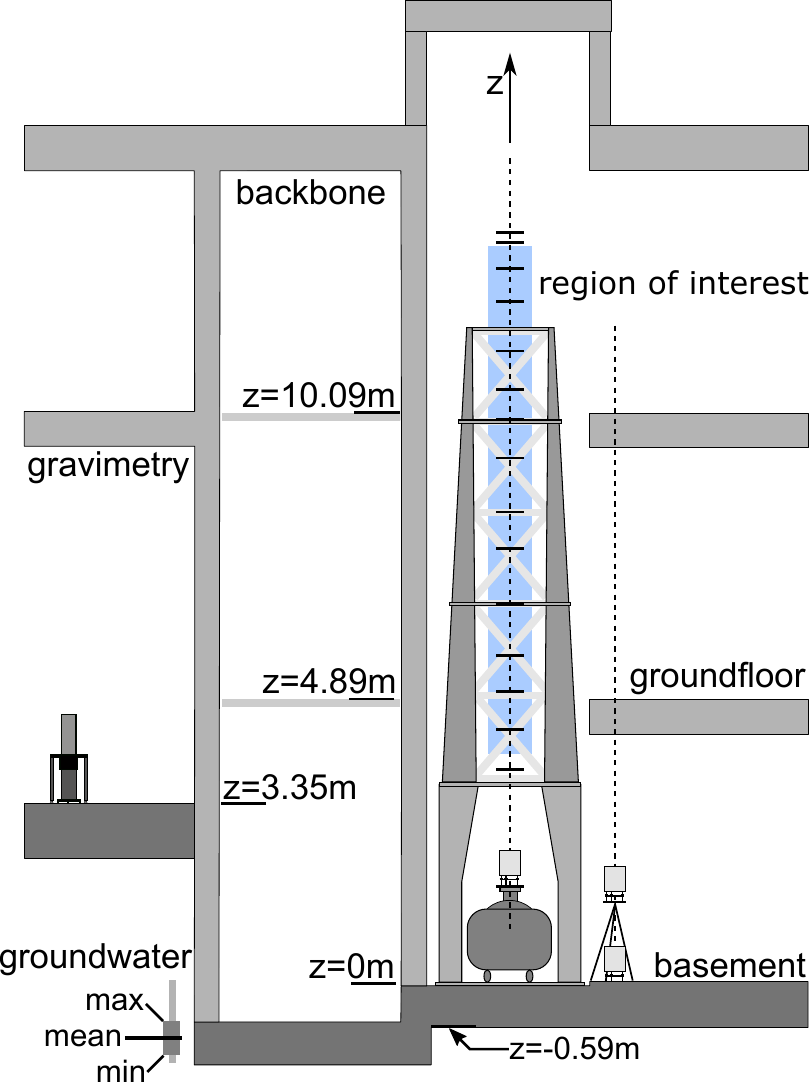}%
    \caption{HITec cross-section (not to scale)}%
    \label{fig:vlbainetz}%
  \end{subfigure}
  \begin{subfigure}[b]{0.59\textwidth}
    \centering
    \includegraphics[height=6.5cm]{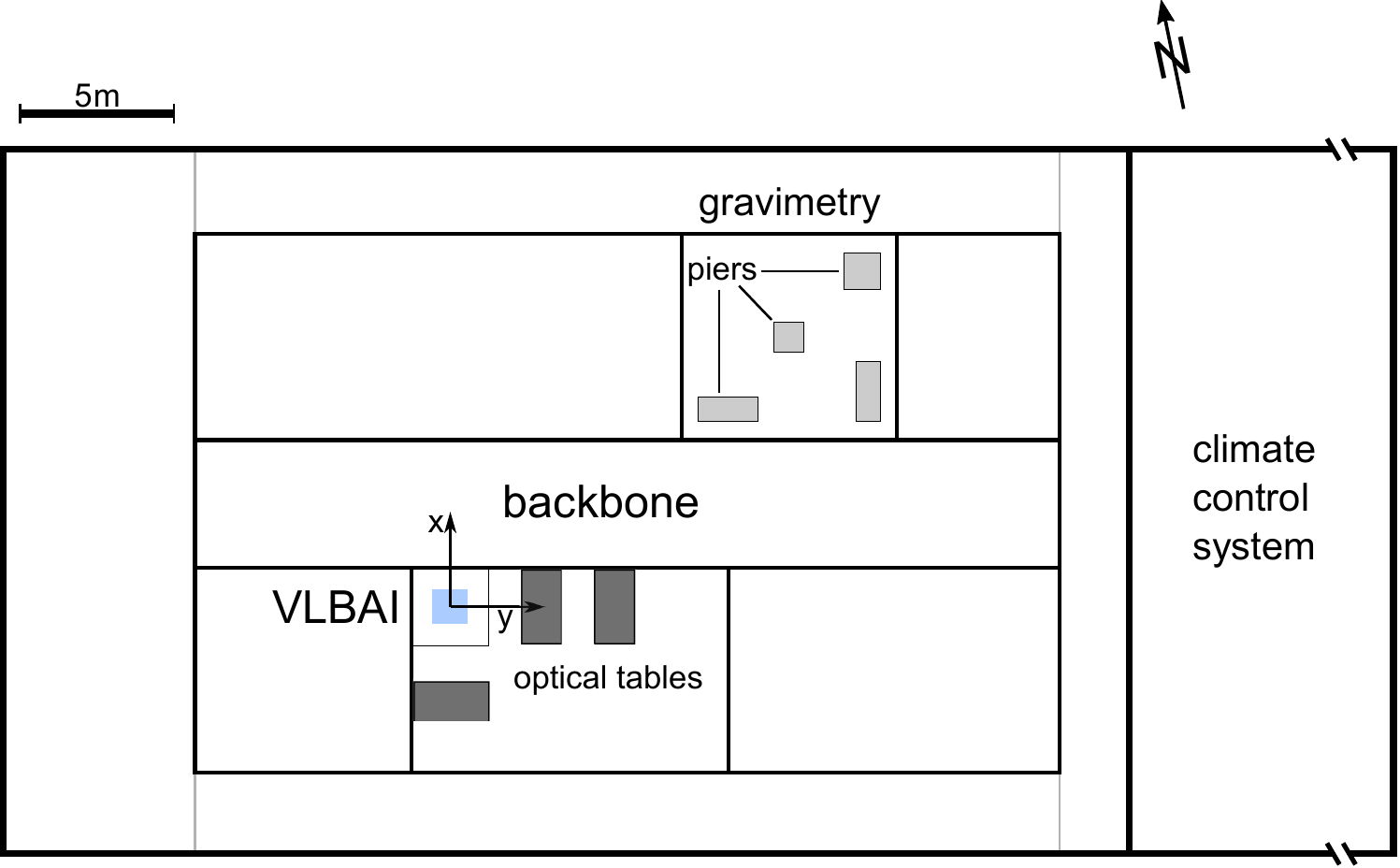}%
    \caption{HITec top view}%
    \label{fig:hitecmap}%
  \end{subfigure}
  \caption{Views of HITec: cross-section (\subref{fig:vlbainetz}) of the VLBAI laboratories with the gravimetric network of 2019 along two vertical profiles and region of interest (blue). The indicated groundwater variation (thick bar) refers to an average annual amplitude of \mwert{0.3}. The thin bar indicates extreme low and high levels. The height $z=\mwert{0}$ refers to the top of the baseplate. The top view of HITec (\subref{fig:hitecmap}) shows the orientation of our coordinate system, the location of the VLBAI facility (blue) and the gravimetry lab including piers for gravimeters (light grey).\label{fig:hitecview}}
\end{figure*}

The VLBAI facility is implemented in the laboratory building of the Hannover Institute of Technology. 
The building consists of three floors (one basement level, two above street level) and is divided into a technical part mainly containing the climate control systems, and a section with the laboratories (see figure~\ref{fig:hitecview}). 
In the laboratory part, a so-called backbone gives laboratories access to the technical infrastructure and divides the building in two parts along its long axis. 
The backbone and southern row of laboratories have a footprint of $\mwert{13.4}\times\mwert{55.4}$ and extend approximately \mwert{5} below surface level.
The northern row of laboratories is fully above ground except for the gravimetry laboratory which is on an intermediate level, around \mwert{1.5} below street level and \mwert{3.4} above basement level (see figure~\ref{fig:vlbainetz}). 
The foundation of the building is \mwert{0.5} thick except beneath the gravimetry laboratory, which has a separate and \mwert{0.8} thick one. 
Figure~\ref{fig:vlbainetz} also shows the measurement points for the relative gravimeters along the VLBAI main axis and a second validation profile, occupied using tripods, next to the VLBAI which were used for the measurements presented in section~\ref{sec:gravmeas}.

\subsection{Physical model}

Following the methods described by \citet{Li1998}, we discretise the HITec building into a model of rectangular prisms that accounts for more than \num{500} elements. 
The geometry is extracted from the construction plans, and we verified all the heights by levelling, also including a benchmark with a known elevation in the German height datum. 
The building is embedded in a sedimentary ground of sand, clay, and marl (\density{2050}). 
For the edifice itself, we include all walls and floors made of reinforced concrete (\density{2500}), the \SIrange{7}{13}{\cm} thick liquid flow screed covering the concrete floors in the labs (\density{2100}), and the gypsum drywalls (\density{800}). 
We also incorporate the insulation material (\density{150}) and gravel on the roof (\density{1350}). 
We use a simplified geometry to model the large research facilities in the surroundings. 
This is for example the case for the Einstein-Elevator \citep{Lotz2018}, a free-fall simulator with a weight of \SI{165}{\tonne} and horizontal distances of \mwert{32} and \mwert{16} to the VLBAI facility and gravimetry laboratory, respectively. 
Finally, we account for laboratory equipment, \eg optical tables (\SI{550}{\kilo\gram} each) according to the configuration at the time of the gravimetric measurement campaigns.

During the first measurements (2017), the interior construction was still in progress, and the laboratories were empty. 
By the time of the second campaign (2019), the building was fully equipped. 
The VLBAI support structure (VSS) and the vacuum tank (VTS) for the seismic attenuation system were in place. 
The VLBAI instrument (atomic sources, magnetic shield, \mwert{10} vacuum tube) and seismic attenuation system were completed after the second campaign. 

Due to their inclined or rounded surfaces, the VLBAI experimental apparatus and its support structure require a more flexible method than rectangular prisms to model their geometry. 
We apply the method described by \citet{Pohanka1988} and divide the surface of the bodies to be modelled into polygonal faces to calculate the gravitational attraction from surface integrals. 
Contrary to the rectangular prisms method, there are only few restrictions on the underlying geometry. 
Most notably, all vertices of a face must lie in one plane and the normal vectors of all surfaces must point outward of the mass. 
For example, normal vectors of faces describing the outside surface of a hollow sphere must point away from the sphere and normal vectors on the inside surface must point towards the centre, away from the mass of the wall of the sphere. 
We extract the geometry of the VLBAI facility components from their tridimensional CAD model through an export in STL\footnote{Stereolithography or standard triangulation language} format \citep{Roscoe1988}. 
This divides the surface of the bodies into triangular faces, therefore ensuring planar faces by default. 
Moreover, the STL format encodes normal vectors pointing away from the object. 
Both prerequisites for the polygonal method by \citet{Pohanka1988} are thus met. 
Using this method, the VSS (aluminium, \density{2650}, total weight \SI{5825}{\kg}) consists of roughly \num{86000} faces and the VTS and corresponding baseplates (stainless steel, \density{8000}, total weight \SI{2810}{\kilo\gram}) contain \num{187000} faces, mostly due to the round shape and fixtures of the VTS. 
As the overall computation time to extract the attraction of these components with a \si{cm}-resolution on both vertical profiles remains in the range of minutes on a desktop PC, we do not need to simplify the models.
The Monte Carlo simulations described in section \ref{sec:transg} nevertheless require the computing cluster of the Leibniz University Hannover (LUH).

We use MATLAB\footnote{MATLAB Version 9.4.0.813654 (R2018a)} to perform the numerical calculations. 
As a cross-check, we implemented both the rectangular prisms and polyhedral bodies methods for the calculation of the attraction effect of the main frame of the HITec building. 
Both approaches agree within floating point numerical accuracy.

\subsection{Time variable gravity changes}
\label{subsec:timegrav}
Mostly for the benefit of the future operations of the VLBAI, we include the effects of groundwater level changes, atmospheric mass change, and Earth's body and ocean tides in our modelling. 
This is necessary for the individual gravimetry experiment (and other physics experiments as well) in the VLBAI on one hand, and for comparing measurements from different epochs, \eg with different groundwater levels, on the other hand.
Previous investigations in the gravimetry lab of a neighbouring building showed a linear coefficient of \nmwert{170} per meter change in the local groundwater table \citep{Timmen2008}. 
This corresponds to a porosity of \SI{>30}{\percent} of the soil \citep{Gitlein2009}. 
For our model, we adapt a porevolume of \SI{30}{\percent}, which has to be verified by gravimetric measurements and correlation with local groundwater measurements.
Two automatic groundwater gauges are available around the building: one installed during the construction work and a second with records dating back several decades also used by \citet{Timmen2008}. 
The effect of atmospheric mass changes is calculated using the ERA5 atmospheric model provided by the European Centre for Medium-Range Weather Forecasts\footnote{https://www.ecmwf.int} and the methods described by \citet{Schilling2019}. 
Tidal parameters are extracted from observational time series \citep{Timmen1994,Schilling2015}. 
Other temporal gravity changes are not in the scope of this work.

Currently, time variable gravity is also monitored with the gPhone-98 gravimeter of the Institute of Geodesy (IfE) at the LUH.
In the long term, we consider the addition of a superconducting gravimeter for this purpose when the VLBAI facility is fully implemented and the experimental work is beginning. 
The support of a superconducting gravimeter is also vital in the characterisation of new gravimeters \citep{Freier2016}.

\subsection{Self-attraction results}
\label{subsec:envmodres}

\begin{figure}%
  \centering
  \includegraphics[width=\columnwidth]{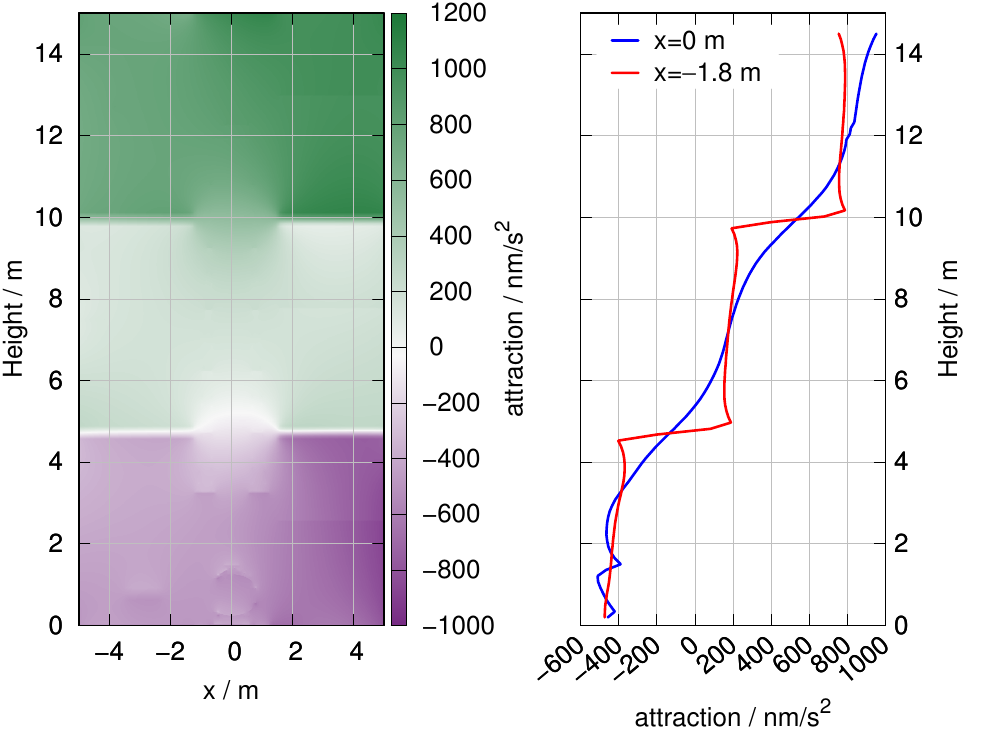}%
  \caption{Calculated gravitational attraction from the building, large laboratory equipment, VSS and VTS in the xz-plane (left) and exemplarily on two profiles (right).}%
  \label{fig:modatt}%
\end{figure}

Figure~\ref{fig:modatt} shows the vertical component of the gravitational acceleration generated by the building, equipment, VSS and VTS. 
The VLBAI main axis is in the centre of the left plot ($x=\mwert{0}$). 
The large structures around \mwert{5} and \mwert{10} correspond to the floor levels. 
Smaller structures are associated to, for example, optical tables or the VSS. 
The right panel of figure \ref{fig:modatt} highlights the attraction calculated for the main axis ($x=\mwert{0}$) and for a second profile along $x=\mwert{-1.8}$ and $y=\mwert{0}$. 
The first profile shows a smooth curve except for the bottom \mwert{2}, which are affected by the VTS. 
In this model, the part above \mwert{2} on the main axis is empty space. 
The second profile, chosen as a sample from the xz-plane, passes through the floors, hence the zig-zag features around \mwert{5} and \mwert{10}. 
While the main axis will later be occupied by the instrument's baseline, this second profile, similar to the validation profile, represents a location that will always remain accessible to gravimeters.

\subsection{Effect of groundwater level changes}
\label{subsec:envgw}
Based on the extensive groundwater level recordings from the gauge nearby the HITec building, we study the impact of groundwater level changes \citep[see also][]{VanCamp2017} on gravitational attraction inside the building, specifically along the VLBAI main and validation profiles, as well as in the gravimetry laboratory.

Due to the layout of the different basement levels in the building (see figure~\ref{fig:vlbainetz}), a change of the groundwater table affects gravity in the VLBAI laboratories differently than in the gravimetry lab. 
Depending on the groundwater level, the foundation beneath the VLBAI laboratories can be partially within the groundwater table, whereas this is never the case for the gravimetry laboratory. 
As shown on figure~\ref{fig:vlbainetz}, the mean groundwater table is nevertheless below the level of the foundation below the VLBAI laboratories. 
Therefore, at certain points of the average annual cycle of amplitude \mwert{0.3}, the groundwater table will rise only around the foundation of the VLBAI laboratories, whereas its level will still increase below the gravimetry laboratory. 
This effect is even more stringent for years where the average cycle amplitude is exceeded (around one in four years).

\begin{figure}%
  \centering
  \includegraphics[width=8cm]{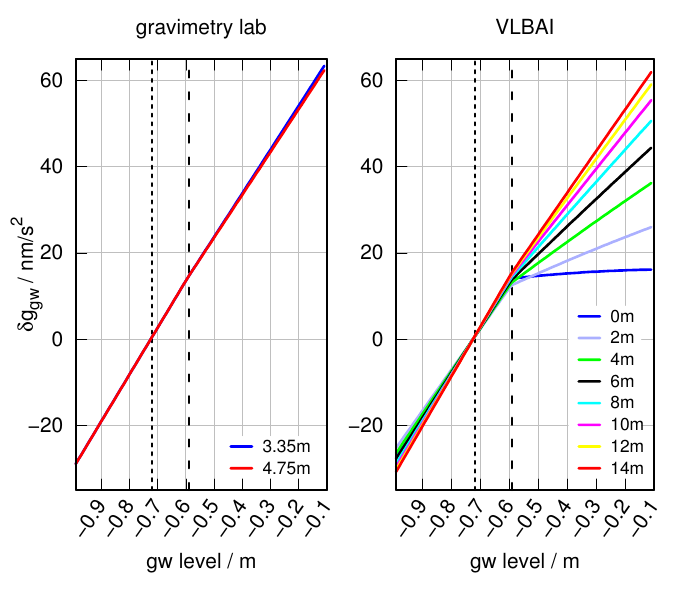}%
  \caption{Effect of groundwater variations (all heights in the height system of the model, cf. figure \ref{fig:vlbainetz}) on gravity in the gravimetry lab (left) and along the VLBAI axis (right) with respect to the mean groundwater level (dotted line \showline{black,dotted}{0.3mm}). 
  The dashed line (\showline{black,dashed}{0.2mm}) indicates the bottom of the foundation below the VLBAI. 
  The coloured lines indicate the change of gravity $\delta g_{\mathrm{gw}}$ in various heights in the gravimetry and VLBAI laboratories. 
  The height of the gravimetry piers in the height system of the model is \mwert{3.35}.\label{fig:gw}}%
\end{figure}

Figure~\ref{fig:gw} illustrates the different influence of the groundwater table level on gravity in the VLBAI and gravimetry laboratories. 
The estimated change of gravity $\delta g_{\mathrm{gw}}$ due to the attraction corresponding to groundwater level variations is presented for different heights above the gravimetry pier and along the VLBAI main axis. 
As the groundwater level is always changing directly beneath the instrument piers in the gravimetry laboratory, we expect an almost linear change of gravity with changing groundwater level. 
The change of gravity is also almost independent of the height above the pier, as shown by the almost identical lines for $z=\mwert{3.35}$ directly on the pier and \mwert{1.4} above the pier, covering the instrumental heights of transportable gravimeters. 
Therefore, AGs with various sensor heights, \eg, A-10 and FG5X, are affected in the same manner. 
The increase of $\delta g_{\mathrm{gw}}$ is \nmwert{32} in an average year. 
This behaviour is different in the VLBAI laboratories. 
In current records, the groundwater level never fell below the foundation of the backbone (cf. figure \ref{fig:vlbainetz}).
This effect is seen in the small divergence (up to \nmwert{3}) for groundwater levels below the foundation of the VLBAI (dashed line).
Once the groundwater level reaches the lower edge of the VLBAI foundation, gravity will not increase linearly along the VLBAI main axis as the groundwater rises further. 
Moreover, in this situation, the effect has a different magnitude depending on the height in the room. 
In a year with the average amplitude of groundwater level variation, ca. \mwert{+-0.15} around the line indicating the mean groundwater level, $\delta g_{\mathrm{gw}}$ will differ by \nmwert{5} between basement and the top floor. 
In years exceeding the average groundwater variation, the difference between the basement and upper levels increases further. 
This effect is within \nmwert{+-2} on the validation profile in the average groundwater cycle.

These observations will be crucial when comparing AGs in the gravimetry laboratory to the VLBAI facility operated as a quantum gravimeter. 
Depending on the geometry of a specific atom interferometer realisation, the instrumental height of the VLBAI gravimeter changes and can introduce changes in the measured value of \textg of more than \nmwert{10} as a result of the groundwater effect in years with a higher than usual groundwater level.
The magnitude of \nmwert{10} is larger than the targeted accuracy of the VLBAI and also a relevant size for classical AGs in comparisons.
It should also be noted, that the model only calculates the gravitational attraction of the groundwater variation. 
A potential vertical displacement of the ground itself is currently not taken into account, leading to a possible underestimation of the effect.

In order to track the effect of groundwater level changes more accurately, we plan to extend the findings of \citet{Timmen2008} by correlating periodic gravimetric measurements on the validation profile in the VLBAI laboratories with the recordings of the two groundwater level gauges around the building. 
This should in particular allow us to take into account that, due to capillarity effects, the groundwater level will probably not sink uniformly below the foundation beneath the VLBAI laboratories once it reaches that level.
\section{Gravimetric measurements}
\label{sec:gravmeas}
In June 2017 and August 2019, we performed surveys using relative gravimeters to verify our model from section~\ref{sec:envmod} along the VLBAI main and validation profiles. 
This approach was already demonstrated in \citet{Schilling2017}, in which the gravity field impact of a \SI{200}{\kilo\newton} force standard machine at the Physikalisch-Technische Bundesanstalt in Braunschweig was modelled. 
That model was verified with gravimetric measurements prior and after the installation of the force machine. 
The difference between the modelled impact and the measurement was within the uncertainty of the gravimeters used. 
For each measurement point, we measured its connection to at least another point and applied the step method with ten connections \citep{Torge2012}. 
A connection corresponds to one gravity difference observation between two points. 
Ten connections require five occupations of a measurement point with a gravimeter. 
We measured most connections with at least two different instruments, reducing the outcomes to a mean instrumental height of \mwert{0.22} above ground or platform. 
We then performed a global least-squares adjustment using the Gravimetry Net Least Squares Adjustment software from IfE \citep[GNLSA,][]{Wenzel1985}. 
The measurements are also calibrated in this process.
We determined the individual calibration factors of the gravimeters on the Vertical Gravimeter Calibration Line in Hannover \citep{Timmen2018,Timmen2020} at least once in the week prior to the measurement campaigns.
The software also corrects Earth tides, applying our observed parameters, and atmospheric mass changes by means of the linear factor of \SI[per-mode=repeated-symbol]{3}{\nms\per\hecto\pascal} with respect to normal air pressure at station elevation. 
In order to account for instrumental drift in the global adjustment, we treat each day and each instrument independently and use a variance component estimation to weight the measurements in the global network adjustment.
The specific groundwater effect discussed in section \ref{subsec:envgw}, considering different magnitudes depending on height, does not apply for either 2017 or 2019 because the groundwater levels were below the foundation of the VLBAI in both years.

\subsection{2017 Gravimetry campaign}
We first mapped the gravity profile along the VLBAI profiles in June 2017, when the HITec building was still under construction and the VLBAI experimental apparatus not yet installed. 
Using the Scintrex CG3M-4492 (short CG3M) and ZLS Burris B-144 (B-114) spring gravimeters \citep{Timmen2004,Schilling2015}, we measured a total of \num{147} connections between seven positions spaced by ca. \mwert{2} along the VLBAI main axis, nine positions on the validation profile, and two points outside of the building. 
We used a scaffolding to access the measurement points on the main axis. 
However, although the scaffold was anchored against the walls, the uppermost platforms were too unstable to ensure reliable measurements. 
The B-114 was only able to measure on the bottom three positions, because the feedback system was not powerful enough to null the oscillating beam on the upper levels. 
The four upper levels were only occupied by the CG3M. 
We connected each point on the scaffold to another one on the same structure and to the closest fixed floor level, at a point part of the validation profile. 
As shown in figure \ref{fig:vlbainetz}, the validation profile included measurements on the floor and on different sized tripods to determine the gradients. 

The variance component estimation gives a posteriori standard deviations for a single gravity tie observation of \nmwert{50} for the B-114 and \nmwert{100} for the CG3M. 
The standard deviations for the adjusted gravity values range from \SIrange{15}{42}{\nms} with a mean value of \nmwert{28}. 
The standard deviations of the adjusted gravity differences vary from \nmwert{21} between fixed floor levels to \nmwert{59} between consecutive levels on the scaffold. 
The transfer of height from the upper floor to the basement through the intermediate levels on the scaffold showed a \SI{2}{\milli\meter} discrepancy compared to the heights from levelling. 
We included the corresponding $\SI{2}{\mm}\cdot\SI[per-mode=repeated-symbol]{3}{\nms\per\mm}=\nmwert{6}$ as a systematic uncertainty for the adjusted gravity values for the values measured on the scaffold. 
We also account for a \SI{1}{\mm} uncertainty on the determination of the relative gravimeter sensor height.

\subsection{2019 Gravimetry campaign}

\begin{figure}%
  \centering
  \includegraphics[width=7cm, trim= 0cm 8cm 0cm 0cm, clip=true]{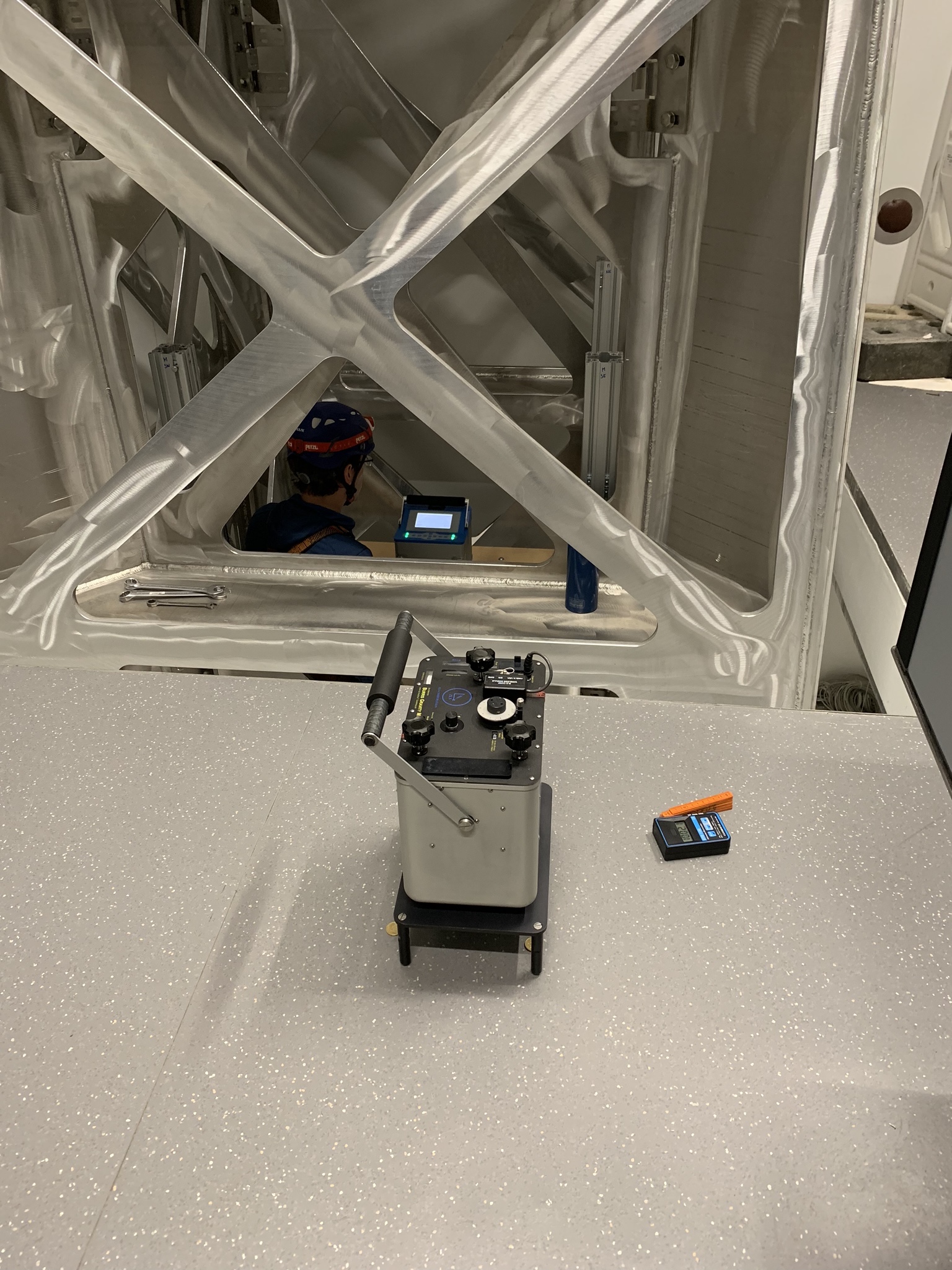}%
  \caption{Measurement at the VSS in 2019 with the B-64 (foreground) on the validation profile and the CG6 (background) inside the VSS on a platform with an operator wearing a security harness. The B-64 is operated on a small tripod to raise the sensor height closer to the CG6 sensor height.\label{fig:meas}}%
\end{figure}

We mapped the gravity profile along the VLBAI axes in a more extensive manner in summer and fall 2019. 
Most measurements were performed in one week of August 2019, adding two days in October and November 2019. 
We used moveable platforms inside the VSS, installed in June 2019, and could measure on \num{16} levels on the main axis, spaced by \SIrange{0.45}{0.95}{\m}. 
The scheme for the validation profile did not change. 
The layout of the network is depicted in figure~\ref{fig:vlbainetz}. 
For this campaign, we used the CG3M, the Scintrex CG6-0171 (CG6), and ZLS Burris B-64 (B-64) spring gravimeters \citep{Timmen2004,Timmen2020,Schilling2015}. 
Owing to the high mechanical stability of the VSS, measurements along the main axis were unproblematic for all instruments and the measurement noise was at a similar level on the moveable platforms and on the fixed floors (see figure \ref{fig:meas}). 
All but one position were occupied with at least two gravimeters, amounting to \num{439} connections in the network adjustment.

The a posteriori standard deviations (single gravity tie measurement) of the observations range from \SIrange{15}{60}{\nms} with more than \SI{50}{\percent} below \nmwert{30}. 
The higher standard deviations are a result of two days of measurements with the CG3M and connections to two particular positions outside of the region of interest of the VLBAI. 
The standard deviations of adjusted gravity values in the network range from \SIrange{7}{19}{\nms} with a mean of \nmwert{9}. 
This improvement, compared to the previous campaign, can be attributed to the stability of the VSS, the addition of the CG6 and the total number of measurements performed. 
The height of the moveable platforms inside the VSS was determined by a combination of levelling and laser distance measurements\footnote{Leica Disto D210} to two fixed platforms and the ceiling. 
For the height determination of the platforms, the uncertainty is \SI{1}{\mm} due to the laser distance measurement. 
We also account for an \SI{1}{\mm} uncertainty in the determination of the instrumental height above the platforms.
\section{Combination of model and measurement}
\label{sec:transg}
\begin{figure*}
	\begin{subfigure}[b]{0.265\textwidth}
		\centering
		\includegraphics[height=7cm]{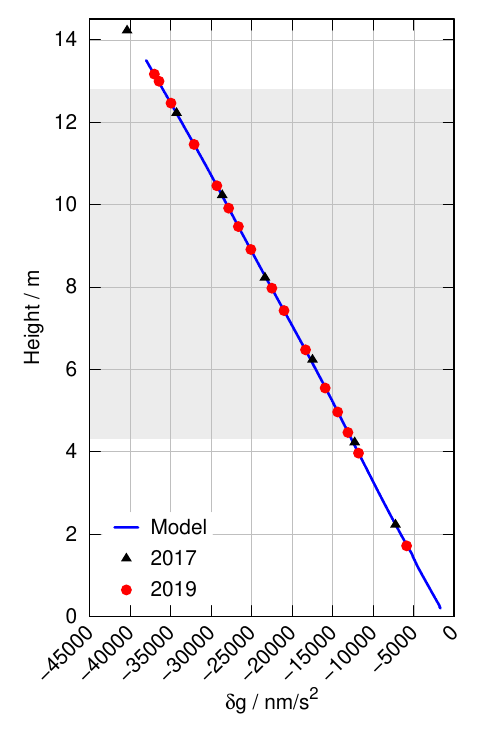}%
		\caption{central axis: gravity variation}%
		\label{fig:vlbaiall1}%
	\end{subfigure}
	\begin{subfigure}[b]{0.24\textwidth}
		\centering
		\includegraphics[height=7cm]{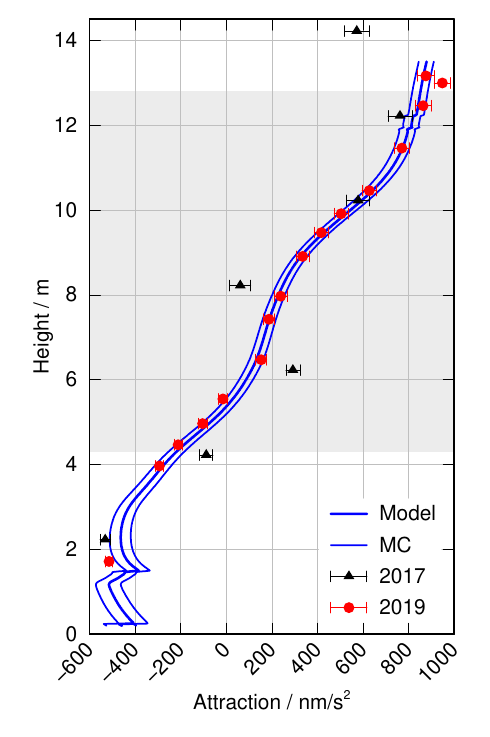}%
		\caption{central axis: model}%
		\label{fig:vlbaiall2}%
	\end{subfigure}
	\begin{subfigure}[b]{0.24\textwidth}
		\centering
		\includegraphics[height=7cm]{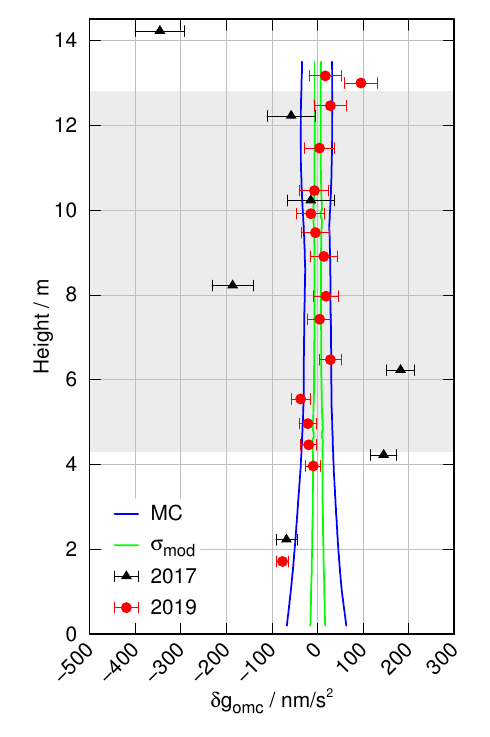}%
		\caption{central axis: residuals}%
		\label{fig:vlbaiall3}%
  \end{subfigure}
	\begin{subfigure}[b]{0.24\textwidth}
		\centering
		\includegraphics[height=7cm]{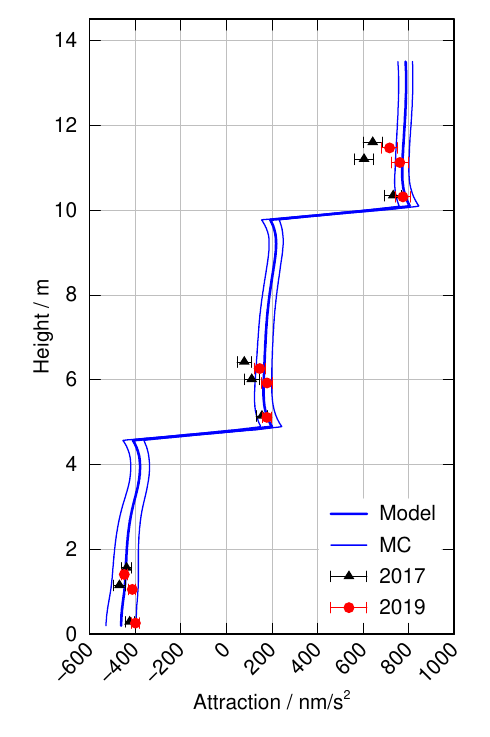}%
		\caption{validation profile: model}%
		\label{fig:vlbaiall4}%
	\end{subfigure}
	\caption{Measurement and model results on the VLBAI central axis (\subref{fig:vlbaiall1}--\subref{fig:vlbaiall3}) and the validation profile (\subref{fig:vlbaiall4}). 
	The shaded area in (\subref{fig:vlbaiall1}--\subref{fig:vlbaiall3}) indicates the region of interest. 
	The total variation of gravity along the central axis is shown in (\subref{fig:vlbaiall1}). 
	The modelled and measured attraction by the environment (with the change of gravity with height removed) on the central and validation profile is shown in (\subref{fig:vlbaiall2}) and (\subref{fig:vlbaiall4}). 
	The errorbars indicate the standard deviations from the network adjustment and the model simulations according to equation \eqref{eq:sigobs}. 
	The maximum and minimum results of the \SI{\pm5}{\percent} density variations from Monte Carlo (MC) simulation of model parameters are indicated by the thin blue lines. 
	The residuals of observations minus model $\delta g_{\mathrm{omc}}$ are given in (\subref{fig:vlbaiall3}) along with the standard deviation of the model $\sigma_\text{mod}$ according to equation \eqref{eq:sigmod}. \label{fig:vlbaiall}}
\end{figure*}

The measurement and model results along the VLBAI main and validation profiles are presented in figure~\ref{fig:vlbaiall}. 
Figure~\ref{fig:vlbaiall1} shows the total variation of gravity along the main axis. 
The plot is dominated by the normal decrease of gravity with height. 
The effect of the building can be better seen when removing the change of gravity with height and visualising only the attraction effect of the building and laboratory equipment, as on figure~\ref{fig:vlbaiall2}. 
There, the model corresponds to the configuration for the 2019 campaign and is identical to the $x=\mwert{0}$, $y=\mwert{0}$ line in figure~\ref{fig:modatt}. 
Figure~\ref{fig:vlbaiall4} shows the model and measurements along the validation profile.

The models presented in figure~\ref{fig:vlbaiall} use the nominal values for the densities of building elements (concrete floors and walls, drywalls, etc.). 
Since these can have variations over the building, we performed a Monte Carlo simulation (\num{50000} runs) varying the densities of the corresponding model elements by \SI{\pm 5}{\percent} according to a normal distribution. 
This leads to a variation of attraction of \SIrange{\pm 27}{\pm 37}{\nms} for heights between \SI{4}{\m} and \SI{13}{\m}, as shown by the thin blue lines on figures~\ref{fig:vlbaiall}\subref{fig:vlbaiall2}--\subref{fig:vlbaiall4}. 
Using a uniform distribution of the density parameters increases the variability by around \nmwert{20}. 
The VSS and VTS are not part of the Monte Carlo simulation since their geometry and materials are well known.

The final location of the VLBAI facility and its main axis could only be approximated to the \si{\centi\meter}-level during the measurement campaigns because of necessary installation tolerances. 
We estimated the effect of a horizontal variation of \SI{\pm3}{\cm} and a vertical variation of \SI{\pm2}{\mm} in a Monte Carlo simulation. 
The total amplitude of the variations at the locations of the gravimetric measurements is within \nmwert{\pm2} with a mean standard deviation of \nmwert{0.3} for the horizontal and \nmwert{0.4} for the vertical component along the main axis.

The measurements, i.\,e. the markers in figure~\ref{fig:vlbaiall}, are the result of the gravity network adjustment. 
Additionally, we removed the effect of the change of gravity with height for figures~\ref{fig:vlbaiall}\subref{fig:vlbaiall2}--\subref{fig:vlbaiall4}. 
For this, the free air gradient is modified with a model of the soil surrounding HITec. 
As the density is only known to a certain degree, the Monte Carlo simulation also included the ground around HITec. 
The standard deviation of the simulation results for each gravimeter position is added to the measurements standard deviation by error propagation. 
The simulations' standard deviations range from \nmwert{10} at the height of \mwert{4} to \nmwert{35} at the topmost position. 
This is also reflected in the increase in the standard deviations indicated by the errorbars in figure \ref{fig:vlbaiall3}.

The uncertainty of the measurements now consists of the following components:
\begin{equation}
	\sigma_{\rm obs}=\sqrt{\sigma_g^2+\sigma_{h,\mathrm{geo}}^2+\sigma_{z,\mathrm{mod}}^2+\sigma_{\mathrm{grad}}^2}\ .
\label{eq:sigobs}
\end{equation}
Here, the standard deviation of the network adjustment is $\sigma_g$. 
The contribution of the determination of the height of the gravimeter is $\sigma_{h,\mathrm{geo}}$. 
The result of the Monte Carlo simulations of the vertical component of geometric position of the central axis $\sigma_{z,\mathrm{mod}}$, and the modelling of the gravity gradient $\sigma_{\mathrm{grad}}$ are also attributed to the measurements.

The standard deviation of the model consists of the following components:
\begin{equation}
	\sigma_{\mathrm{mod}}=\sqrt{\sigma_{\mathrm{MC}}^2+\sigma_{hz,\mathrm{mod}}^2}\ ,
\label{eq:sigmod}
\end{equation}
where $\sigma_{\mathrm{MC}}$ is the standard deviation of the Monte Carlo simulations of the model density, calculated in the heights of the gravimetric measurements, and $\sigma_{hz,\mathrm{mod}}$ is the standard deviation of the Monte Carlo simulations for the horizontal component of the geometric positions along the VLBAI main axis. 
$\sigma_{\mathrm{mod}}$ is shown in figure \ref{fig:vlbaiall3} with a range of \SIrange{6}{11}{\nms} in the region of interest and about $\nmwert{8}$ at $\zeff = \SI{9.2}{\meter}$ (see section~\ref{subsec:effh}).

Furthermore, a single parameter is estimated to reduce the gravity values from the magnitude of \SI{9.81}{\mps} to the order of magnitude of the model values for the attraction. 
This parameter is the mean difference of observed minus computed results at the location of the observation in the region of interest. 
The measurements of 2017 are also corrected for the changes within the building with respect to 2019. 
No additional parameters were estimated to fit the measurements to the model or vice versa. 
The remaining signal should now contain the effect of the HITec building on gravity.

In general, the 2017 measurements and the main axis model do not show a good agreement \citep[see also][]{Schilling2019} due to the instability of the scaffolding used as a platform \citep[see also][]{Greco2014}. 
The agreement on the validation profile is better, and only the two topmost points do not agree with the model and simulation. 
These earlier measurements serve as a proof of concept and are given for the sake of completeness. 
The following discussion concerns only the 2019 measurements.

The 2019 campaign provides a clear improvement considering the number of stations along the VLBAI main axis, the stability of the platforms in the VSS and therefore data quality. 
Consequently, the agreement between measurement and model is significantly improved. 
The measurement scheme on the validation profile remained unchanged compared to the 2017 campaign. 
Figure~\ref{fig:vlbaiall3} shows the difference between the measurements and the model on the central axis.
The region of interest for experiments in the VLBAI is approximately between \mwert{4} and \mwert{13} (see figure~\ref{fig:vlbai}). 
Within this region, only the second-highest point is not within the simulation's \SI{\pm 5}{\percent} density variations. 
The two tailed statistical test ($\alpha=0.05$) on the equality of model $\delta g_{\mathrm{mod},i}$ and measurement $\delta g_{\mathrm{obs},i}$ at point $i$ according to
\begin{align*}
	&\text{Null hypothesis:} & \delta g_{\mathrm{omc},i}=&\delta g_{\mathrm{obs},i}-\delta g_{\mathrm{mod},i}=0\\
	&\text{Alternative hypothesis:} & \delta g_{\mathrm{omc},i}\neq &0\\
	&\text{Test statistics:} & t_i=&\frac{\left|\delta g_{\mathrm{omc},i}\right|}{\sqrt{\sigma_{\mathrm{obs},i}^2+\sigma_{\mathrm{mod},i}^2}}
\end{align*}

passes for all but three points. 
The null hypothesis, considering the symmetry of the normal distribution, is rejected if $t_i>N_{(0,1,1-\nicefrac{\alpha}{2})}$. 
The test fails for the points at $z=\SIlist[list-final-separator = {\text{\ and\ }}]{1.72;5.55;12.99}{\m}$.

The lowest point at $z=\mwert{1.72}$, directly on the VTS, was challenging to measure, as the pump of the vacuum tank was active during the measurements causing high-frequency vibrations. 
As this position is outside of the experimental region of interest, no additional measurements were taken. 
The cause for the significant deviation from the model at $z=\mwert{12.99}$, which was measured with only one gravimeter, is unknown. 
The height difference to the point above is only \mwert{0.16} of free space, so a real gravity variation appears unlikely. 
Treating this point as an outlier, and repeating the test after calculating the offset between adjusted gravity values and model without this measurement, the test also passes for the point at $z=\mwert{5.55}$. 
All points on the validation profile pass the statistical test. 
The standard deviation of observations minus model is \nmwert{20} (\nmwert{31} if the second-highest point is included) for the central axis in the region of interest and \nmwert{34} on the validation profile.

The density of the different model components, chosen initially from technical documentation, are sufficient to generate a model which is identical to in situ measurements at a \SI{95}{\percent} confidence level. 
Modelling a \SI{5}{\percent} normally distributed variation of these densities results in a narrow range of possible model variations, which covers almost all measurements used to verify the model. 
We expect that using individual densities for each floor instead of one common density value for all concrete components in the building would improve the agreement between model and observations on the validation profile. 
Such extra modelling step should however be constrained not to deteriorate the model accuracy in the experimental region of interest.

As a final step, the VLBAI magnetic shield and vacuum system \citep[][installed December 2019]{Wodey2019} will be added to the model. 
Similarly to the VSS and VTS, this component was designed using CAD, built with known materials, and can be exported into the required format for our model. 
While the assembly is significantly more complex, we expect the octagonal symmetry of the magnetic shield to simplify the numerical calculations and allow us to reach the same level of accuracy in the gravity model as for the VSS and VTS. 
It will however only be possible to check the quality of the extended model with measurements on the validation profile, as the main axis is obstructed by the instrument's vacuum chamber. 
Nevertheless, the understanding of environmental variations (mostly hydrology) outlined in section~\ref{subsec:envgw} will render this possible with good accuracy. 
Due to the work associated with the installation of the VLBAI baseline components, this last model extension and its corresponding validation have not been done yet.

Extending our model with the VLBAI baseline components will allow us to connect gravimetric measurements along the validation profile and future data acquired by a VLBAI quantum gravimeter along its main axis in our adjusted gravimetric network. 
Since the measurement positions along the validation profile will remain free during operation of the VLBAI facility, this will for example enable comparisons of the VLBAI QG with FG5(X)-type classical AGs positioned in the VLBAI laboratories. 
In this specific setup, contributions of time variable gravity to the measurements are minimal for the VLBAI and instrument under test. 
To further minimize the height dependency due to the groundwater effect, the atom interferometer could be realised with an effective height close to the instrumental height of the classical AG, \eg with the AG on the groundfloor.
Taking into consideration the mean standard deviation of the relative gravimeter network of \nmwert{9}, we expect to be able to transfer \textg with an uncertainty of \nmwert{10} and possibly below from the VLBAI baseline. 
Furthermore, creating a similar network including stations along the validation profile and in the HITec gravimetry laboratory would permit gravimetric comparisons between the VLBAI QG and instruments operated on the gravimetric piers.
The estimates so far exclude the inevitable contribution of the VLBAI gravity measurement.
The determination and validation of the VLBAI uncertainty budget will be published in a separate study.
\section{Conclusions}
We established a gravimetric control network for the Hannover VLBAI facility, a novel \mwert{10}-scale atom interferometer. 
The network consists of \num{439} connections measured by relative gravimeters. 
A least squares adjustment of the network results in a mean standard deviation of the adjusted gravity values of \nmwert{9}. 
In addition, we developed a structural model of the building hosting the VLBAI facility and its surroundings. 
When compared, the model and the measurements agree with \SI{95}{\percent} confidence, with standard deviations of the residuals of \nmwert{20} along the atom interferometer's baseline, and \nmwert{34} on a second, parallel profile. 
Moreover, we gained insight on some dynamical aspects of the gravity field around the instrument, namely the effect of groundwater level variations. 

We anticipate this gravimetric network to contribute to the assessment of the quantum gravimeter's uncertainty budget, which is currently not included in our study.
The current work is also essential to help determining the effective instrumental height (\textg-value reference position) and enable transfers of \textg values from the atom interferometer's baseline to the validation profile, accessible to mobile gravimeters for comparison and possibly calibration purposes, at the \nmwert{10} repeatability level (relative to the VLBAI deduced g-values). 
Completing the model by including the VLBAI baseline, refining the description of the soil surrounding the host building, and including better estimates for the building material densities, we expect to shift the possibility for gravity field measurement transfers and mobile instrument calibration towards the \nmwert{5} level, improving the temporal stability of the current state of the art, which is still largely based on gravimeter comparisons.
This paves the way for the realisation of a new gravity standard based on atom interferometry. 
Finally, the knowledge of the dynamical gravity field and its gradients is key to reaching new frontiers in fundamental physics tests with very long baseline atom interferometry.
{\small
\begin{acknowledgements}
  The Hannover Very Long Baseline Atom Interferometry facility is a major research equipment funded by the Deutsche Forschungsgemeinschaft (DFG, German Research Foundation). This work was supported by the DFG Collaborative Research Center 1128 ``geo-Q'' (project A02, Contract Number 239994235) and is supported by the CRC 1227 ``DQ-mat'' (project B07, Contract Number 274200144), Germany\textquotesingle s Excellence Strategy -- EXC-2123 ``QuantumFrontiers'' -- 390837967, and the computing cluster of the Leibniz University Hannover under patronage of the Lower Saxony Ministry of Science and Culture (MWK) and the DFG.

  M.~S., \'E.~W., and C.~S. acknowledge support from ``Nieders\"achsisches Vorab'' through the ``Quantum- and Nano-Metrology (QUANOMET)'' initiative (project QT3), and for initial funding of research in the DLR-SI institute.  D.~S. acknowledges funding from the German Federal Ministry of Education and Research (BMBF) through the funding program Photonics Research Germany (contract number 13N14875).

  The VLBAI support structure was conceived by the engineering office Heinz Berlin (Wennigsen, Germany) in collaboration with the VLBAI science team, and produced by Aljo Aluminium-Bau Jonuscheit GmbH (Berne, Germany).

  We thank W. Ertmer for his vision and long lasting support on very long baseline atom interferometry and the acquisition of funding for the Hannover Institute of Technology. We are grateful to T. Frob\"ose and A. Wanner for their assistance during the installation of the vacuum tank and support structure.

  We thank the three reviewers for their valuable input to improve this article.
\end{acknowledgements}
\begin{authorcontrib}
M.S., \'E.W., L.T. planned geometric and gravimetric measurements, evaluated the data and prepared the initial draft.  \'E.W., D.T., D.S., C.S., E.M.R. conceptualised VSS, VTS. \'E.W., D.T., K.H.Z. designed and built measurement platforms for VSS. M.S., \'E.W., L.T., D.T., K.H.Z. carried out the measurements. M.S. developed and implemented the gravity model. D.T., K.H.Z., D.S., C.S., E.M.R., J.M. provided critical input to the manuscript and approved the final version.
\end{authorcontrib}
\begin{dataavail}
 Data of absolute gravimeter key comparisons is available in the Key Comparison Database (https://www.bipm.org/kcdb) and cited literature. Gravimetric measurements in instrument specific ascii data formats and datasets generated in this study are available from the corresponding author on reasonable request.
\end{dataavail}
}
\printbibliography
\end{document}